\documentclass{IEEEtran}
\usepackage{cite}
\usepackage{amsmath,amssymb,amsfonts}
\usepackage{algorithmic}
\usepackage{graphicx}
\usepackage{textcomp}
\usepackage{url}
\usepackage{multirow}
\usepackage{enumerate}
\def\BibTeX{{\rm B\kern-.05em{\sc i\kern-.025em b}\kern-.08em
    T\kern-.1667em\lower.7ex\hbox{E}\kern-.125emX}}
\begin{document}
\title{Review of Serious Games for Medical Operation}
\author{Huansheng Ning, \IEEEmembership{Senior Member, IEEE}, Zhijie Guo, Raouf Hamzaoui, \IEEEmembership{Senior Member, IEEE}, Rongyang Li, Fadi Farha, and Lingfeng Mao
\thanks{This work was supported by the National Natural Science Foundation of China (61872038), and in part by the Fundamental Research Funds for the Central Universities under Grant FRF-GF-19-020B. (Corresponding author: Huansheng Ning.)}
\thanks{Huansheng Ning is with the School of Computer and Communication Engineering, University of Science and Technology Beijing, Beijing 100083, China, and also with the Beijing Engineering Research Center for Cyberspace Data Analysis and Applications, Beijing 100083, China (e-mail: ninghuansheng@ustb.edu.cn). }
\thanks{Zhijie Guo, Rongyang Li, Fadi Farha and Lingfeng Mao are with the School of Computer and Communication Engineering, University of Science and Technology Beijing, Beijing 100083, China}
\thanks{Raouf Hamzaoui is with the School of Engineering and Sustainable Development, De Montfort University, The Gateway, Leicester, LE1 9BH}}

\maketitle

\begin{abstract}
\textit{Objective}: Medical operations(MOs) are essential in healthcare, and they are also a big concept that includes various operations during the perioperative period. Traditional operation exposes its limitations during the perioperative period, reflected in medical training, surgical preparation, and postoperative rehabilitation. Serious Games for Medical Operation (SGMO) offer new ways and complementary solutions to support MOs. \textit{Methods}: As a review, this paper analyzes the development of SGMO and considers various aspects of the SGMO, such as interface, functions, and technologies. \textit{Results and conclusion}: By combining MO and serious games' characteristics, the paper classifies SGMO and analyzes their features and functions for different groups of users and at various stages of the perioperative period (before, during, and after an MO). Interactive technologies used in SGMO are presented from a visual, haptic, and auditory perspective. \textit{Significance}: This paper reviews the development of SGMO, summarizes its functions and technologies. Besides, it presents representative products and suggests future research directions.
\end{abstract}


\section{Introduction}
\label{sec:introduction}
Surgery refers to the process of treating injuries and diseases by removing, repairing, or readjusting organs and tissues, often involving cutting into the body\cite{surgery}. Since the development of medicine, surgery has always been one of the most important disciplines. However, in 2015, nine of ten people in low-income and middle-income countries did not have access to safe and affordable surgical care\cite{2017Global}. In this paper, a medical operation (MO) refers to all operations in the perioperative period, whether surgical or nonsurgical. Although the traditional MOs have greatly benefited humankind, their limitations should not be underestimated. In the perioperative period, MOs are constrained by time and space, and the cost of human and material resources is high\cite{9107339,4518990,6212340}. For this reason, researchers have been looking for new technologies to innovate MOs.\indent\par

In particular, many simulators have been developed and successfully used in  surgical procedures. For example, Zhang et al.\cite{4407363} proposed a full-fledged MIS simulator to study tumor resection. Omata et al.\cite{8266126} established an eye surgery simulator for training of peeling the inner limited membrane. Li et al.\cite{9089445} used virtual reality laparoscopic simulators to enhance the reality of surgical training. Serious games (SGs), as games not designed for entertainment, mainly solve problems by simulating real events or processes. The use of serious games in medical operations has attracted increasing attention in recent years. Mert et al.\cite{8493405} analyzed the game's hierarchical structure and content. Tan et al.\cite{8056599} collected and reviewed research on nursing education in SGMO. Graafland et al.\cite{graafland2014systematically} proposed a consensus-based framework for the assessment of specific medical serious games. Baby et al.\cite{7724725} reviewed SGMO from an educational and learning point of view.  However, there is no overview of the development of SGMO. Besides, various classification methods lead to unclear research directions, which is not convenient for researchers. To fill the gaps in this research, we gave a clear overview of the development of SGMO.\indent\par

SGMO have a wide variety of functions, and users choose the appropriate SGs based on the required functions. According to the different groups corresponding to different functions of SGMO, we divide SGMO into two categories. The first category is applied to medical staff, and the second category is applied to patients undergoing MOs. The medical staff is the operational process's performer and represents the active party in MO. SGMO need to provide training and testing functions to improve the medical staff's skills in the perioperative period\cite{8056599},\cite{7724725,6627863,8401365}. On the other hand, the patients are the passive party in the MOs and have different physical and psychological conditions. In the perioperative period, a series of procedures such as diagnosis, cooperation, treatment, and rehabilitation are required for patients, so SGMO should meet these needs and provide different help to patients at different periods\cite{pascual2012diagnosis,yao2018development,vilacca2019diagnosis}. Interactive technology is an essential technical component in SGMO\cite{marshall2015serious},\cite{overtoom2019haptic}. In MOs, medical staff need to notice the details of the patient's operation and various physiological indicators with multiple senses, the most important being vision, touch, and hearing\cite{mack2012interactive}. No previous paper provided a comprehensive analysis of SGMO technology from a sensory interaction perspective. This paper fills this gap by analyzing the interactive technology in SGMO from three sensory aspects and presents representative products and SGs.\indent\par

The remainder of the paper is organized as follows. Section II presents the development process of SGMO. Section III classifies SGMO according to different groups of users. Section IV reviews the interactive technology in SGMO from different sensory perspectives. Section V lists future research directions of SGMO. Finally, Section VI draws the conclusion.

\section{Development of Serious games in medical operation}
\label{sec:development}
SGMO are developing rapidly because they have a great potential and effective functions in addressing the limitations of traditional MOs\cite{8401365}. At the same time, SGMO are of interest to both computer scientists and medical doctors. To analyze SGMO effectively, it is important to consider various aspects of the game, such as interface, technology and functions.\indent\par

The early form of serious games was board games. Participants in board games meet face to face to discuss based on the games’ content. Some patients with dementia will increase their interest in communicating with others during the games, and tangible props will also increase their hands' flexibility\cite{ning2020review}. Whittam et al.\cite{whittam2017educational} used a board game called Burns in their research, which contains a folding board. Many question cards are mainly divided into seven categories about pathological knowledge and treatment methods for burn wounds. The final study involving the medical staff who participated in this game showed that using Burns is an easy learning process. Although board games have a simple format and do not use advanced technology to train medical staff and help patients improve their condition, they are of great significance in the early development of SGMO.\indent\par

Video games usually contain the basic control equipment (such as computers, pads, mobile phones, mice, keyboards, etc.). They are shown on electronic screens and add video and audio technology to enrich their functions. Video games can show realistic surgical scenes, allowing medical staff to observe the patient's condition. Sabri et al.\cite{sabri2010off} used a first-person-shooter game to show the surgical environment in Off-Pump Coronary Artery Bypass Surgery (OPCAB). The model rendered by the game is very accurate, and the lighting effects are the same as in the operating room. Lee et al.\cite{lee1999interactive} used video games to show the shape and structure of the colon's inner surface. Moreover, the navigation screen in the game is simple and clear. Graafland et al.\cite{graafland2014serious} used a video game that can be run directly on mobile phones to simulate biliary surgery and allow players to make clinical decisions. The blood vessels and tissues are visible on the screen and presented during the game.\indent\par

Somatosensory games have become more complicated. In addition to the game itself and basic control equipment, this type of game adds joysticks or sensing devices. These games' technology is also more complicated, and the body position can be input or captured through peripheral devices. Carbone et al.\cite{carbone2016psychomotor} provided laparoscopic surgeons with gamepads to move virtual objects by manipulating the handles. The game's opponent movements are required to be very precise, just like a real laparoscopic surgery. Piedra et al.\cite{7590375} used Leap Motion to capture gestures and control the character model to complete the tasks. Ryden\cite{6193304} used Kinect to capture human limbs' movements and make the whole body become a controller. Jawaid et al.\cite{8250032} verified that Kinect has a better recognition effect in diagnosing the body. Since somatosensory games can capture the posture and movements of the human body through peripheral devices, they make the field of remote MO a reality. Guthart et al.\cite{844121} studied the daVinci™ System, which consists of two sub-systems: a surgical console for doctors and an operating cart for patients. The doctor transmits position, direction, and grasping commands by operating the handle. The patient-side tool has multiple degrees of freedom and can be connected to various surgical instruments and the tool complete the operation according to the doctor's actions. Noonan et al.\cite{4651105} added an eye-tracking device to the DaVinci system, and the experiment proved that it had a real-time control effect. Ruszkowski et al.\cite{7139812} used the DaVinci system to evaluate surgical skills in coronary artery bypass surgery. Although there are no special SGs for remote MOs, the equipment and functions provided by somatosensory games are very suitable for remote MOs, which significantly enriches the types of SGMO.\indent\par

Virtual games, typically represented by Virtual Reality (VR), Augmented Reality (AR), and Mixed Reality (MR) games, have developed rapidly in recent years. This type of games has many forms and powerful functions. It uses advanced virtual technology to make players feel immersed in the game world. Correspondingly, game development and costs are high. For example, Surgeon Simulator: Experience Reality is a VR serious game released in 2016, which requires a VR device to experience the game. The player is assigned the role of a doctor to perform surgery on the patient. By moving the head freely, the player can see different perspectives from the VR glasses and experience a realistic surgery environment. Larrarte and Alban\cite{7743367} used AR games on a tablet to help doctors view additional information about surgery. The game recognizes various tags to present virtual images and trains for minimally invasive surgery.\indent\par

Compared with other types of games, virtual games have more advanced technology and more powerful functions. The development of SGMO and the comparison of different types of games are shown in TABLE I.

\begin{table}[]
	\centering
	\label{Table1}
	\setlength{\tabcolsep}{3pt} 
	\renewcommand\arraystretch{1.5} 
	\caption{DEVELOPMENT AND COMPARISON OF DIFFERENT TYPES OF SGMO}
	\begin{tabular}{|c|c|c|}
		\hline
		Users                          & Classification & Focused functions                                                                                                      \\ \hline
		\multirow{2}{*}{Medical staff} & Training       & \begin{tabular}[c]{@{}c@{}}Provides training in accordance\\  with expertise\end{tabular}                              \\ \cline{2-3} 
		& Testing        & \begin{tabular}[c]{@{}c@{}}Gives a comprehensive judgment by\\  combining scores and review\end{tabular}               \\ \hline
		\multirow{3}{*}{Patients}      & Before MO      & \begin{tabular}[c]{@{}c@{}}Eliminates preoperative anxiety and panic\\  and improves surgical cooperation\end{tabular} \\ \cline{2-3} 
		& During MO      & \begin{tabular}[c]{@{}c@{}}Detects whether functional neurons in the\\  brain are damaged\end{tabular}                 \\ \cline{2-3} 
		& After MO       & \begin{tabular}[c]{@{}c@{}}Provides postoperative rehabilitation\\  and reduces postoperative pain\end{tabular}        \\ \hline
	\end{tabular}
\end{table}

\section{Serious games for different groups of people in medical operation}
\label{sec:functions}

\begin{table*}[]
	\centering
	\label{Table2}
	\setlength{\tabcolsep}{3pt} 
	\renewcommand\arraystretch{2.0} 
	\caption{SGMO’s DIFFERENT FUNCTIONS FOR DIFFEENT GROUPS}
	\begin{tabular}{|c|c|c|}
		\hline
		Users                          & Classification & Focused functions                                                           \\ \hline
		\multirow{2}{*}{Medical staff} & Training       & Provides training in accordance with expertise                              \\ \cline{2-3} 
		& Testing        & Gives a comprehensive judgment by combining scores and review               \\ \hline
		\multirow{3}{*}{Patients}      & Before MO      & Eliminates preoperative anxiety and panic and improves surgical cooperation \\ \cline{2-3} 
		& During MO      & Detects whether functional neurons in the brain are damaged                 \\ \cline{2-3} 
		& After MO       & Provides postoperative rehabilitation and reduces postoperative pain        \\ \hline
	\end{tabular}
\end{table*}

SGMO must mirror a real MO. Two main groups are involved in MOs : medical staff and patients, as the active and passive sides of the operation. We first clarify when these two groups of people use SGMO, considering that their requirements  are generally different. Therefore, this section summarizes the different functions of SGMO from the perspective of different users. TABLE II shows the relationship between functions and groups.


\subsection{SGMO for medical staff}
Medical staff lead the entire operation directly, receive training before the operation, and review the patients condition after the operation. A medical operation is also complex, and the medical staff’s attention is required at all time as minor mistakes can have dire consequences. Therefore, SGMO cannot be used during the operation, but they can train the medical staff and test their professional abilities. When medical staff use SGMO, they do not execute an operation but use the functions provided by SGMO to receive corresponding training to obtain declarative knowledge and procedural knowledge\cite{2019Serious}. At the same time, SGMO should also be used as a tool to test training effects\cite{graafland2014serious}. In summary, SGMO are mainly divided into two categories for medical staff: training category and testing category.

\subsubsection{Training}
Training serious games’ functions are mainly embodied in the preoperative training, and the training process aims to learn the relevant knowledge of MO. The medical staff have various learning methods in this learning process, such as traditional lectures, audio teaching, and video teaching. The purpose of these teaching methods is to provide theoretical knowledge related to MO\cite{2019Serious}. These theoretical knowledge systems are so complicated that the teaching cycle is long and teaching tasks are difficult. Besides, some professional knowledge explanations may require specific medical environments and surgical instruments, so the limited medical resources may not meet the teaching goals.\indent\par

Training medical staff with SGMO is different from training them with traditional teaching methods. Different SGMO users have different theoretical knowledge and skills, but the aim is to provide declarative knowledge and help medical staff convert it into procedural knowledge. Tan et al.\cite{8056599} proposed SGs as a teaching tool which can enhance learning motivation and encourage learners to participate in the learning process actively. Similarly, Byl et al.\cite{8401365} used a VR game, in which the environment is  a toy factory. Subjects were asked to use spatial imagination and graphic stitching according to the game requirements to simulate the operational process. The results showed that the participants generally believed that the game is attractive and motivating. Another finding was that the training effect was improved by allowing subjects to use the game multiple times. Chon et al.\cite{2019Serious} pointed out that the use  of SGMO can significantly improve the declarative knowledge and procedural knowledge of medical staff. In this study, a VR game called EMERGE was used with good results. IJgosse et al.\cite{2018Saving} used a video game called Underground that mainly demonstrated the transfer of serious games to laparoscopic simulator technology and analyzed that if SGs can be transferred to the operating room, it will be a valuable and cost-effective additional contribution to MO. Similarly, in minimally invasive surgery, Piedra et al.\cite{7590375} used a VR serious game that simulated the real surgical environment, requiring the subject's hand movements to complete the operation. The results showed that SGMO could improve hand-eye coordination. In addition to knowledge training, the learning process for medical staff can also call attention to medical equipment-related issues. Equipment and instrument failures occasionally occur, but traditional training often neglects them. Graafland et al.\cite{2017Game} mainly trained subjects to deal with equipment-related failures and finally showed that handling them should also be part of the training and can reduce the corresponding risks during the operation. When doctors diagnose and treat patients, they often need to use their experience to determine the severity of the patient's trauma and decide whether to triage. Mohan et al.\cite{2018Serious} used a video game to improve doctors' heuristic judgment ability. Randomized controlled trials showed that games have a positive effect on improving the ability, while traditional education has no effect. It can be seen that SGMO are a new solution that can increase the enthusiasm of medical staff to receive training, improve the effectiveness of learning, address the lack of medical resources, and improve the heuristic judgment ability.

\subsubsection{Testing}
The function of this type of games is to assess the medical staff's knowledge level and professional skills. Compared with traditional assessment methods (such as exams and skill tests), it may be a new and effective form. The examination is a typical testing method to verify whether the medical staff has mastered knowledge during training. No matter how detailed the exam is, it cannot cover all aspects of  medical knowledge. Piedra et al.\cite{7590375} designed a VR game to evaluate the doctors' proficiency in surgical skills. There were three steps in this game consistent with the cutting and suturing in a MO. The correctness, completeness, and time used to complete the operation are used as judging criteria and counted as a score. The score in this system reflects the skills, and the evaluation has three levels: easy, medium, and difficult, which are appropriate for doctors with different qualifications. Graafland et al.\cite{graafland2014serious} used a video game called Medialis to test the doctor's clinical decision-making ability. This game randomly generates multiple cases and allows the player to make a diagnosis. The time for each case should not exceed ten seconds. After each diagnosis, the player ges a score, which can not only objectively evaluate the accuracy of the clinical decision-making but can also be uploaded to social platforms to enhance the popularity and enthusiasm. De Lima et al.\cite{2016A} used a virtual game to test the professional level of medical students. The game also tests the computer system through dialogue with Non-Player Characters (NPCs), combining SGMO with machine learning. It can be seen that testing games can be used as a new method to evaluate the professional skills of medical staff and have great development potential.

\subsection{SGMO for patients}
The previous section mentioned that SGMO can be used as a training or testing tool for medical staff without being directly used in the surgical process. Patients use SGMO to cure physical or psychological diseases to recover as soon as possible. This includes preoperative preparations, surgical procedures, and postoperative rehabilitation. Therefore, SGMO can be used in the entire process of the real operation for the patient and provide different kinds of assistance to the patient at different stages of the operation. Because of the functional requirements of patients for SGMO throughout the entire procedure, it is useful to study the functions of SGMO chronologically. Besides, SGMO have various functions for patients, such as improved surgical cooperation, preoperative sedation, and postoperative rehabilitation. It is difficult to divide such functions into several categories. Therefore, we mainly analyze the functions of SGMO applicable to patients from the  preoperative, intraoperative, and postoperative perspectives.

\subsubsection{Using SGMO before a medical operation}
When patients are waiting for an operation, they are often affected by psychological burdens and feel anxiety in many cases. For example, some children may be restless and refuse to cooperate. Drummond et al.\cite{2018} showed how a video game named Staying Alive could familiarize children with the hospital environment and surgical procedures, eliminating their fear of surgery. Sourina et al.\cite{PMID:21335865} regarded SGs as a new method to supplement traditional drug treatment and created an EEG-based game for pain management. Buffel et al.\cite{Buffel2019A} focused on the anxiety and pain of children before the operation. They provided a serious game called CliniPup to the children in a test group as a functional training before operation. They used the modified Yale Preoperative Anxiety Scale to measure the children's anxiety and the Wong-Baker Faces Pain Rating Scale to test the children’s pain after the operation. The results showed that although there was no significant difference in children's postoperative pain scores, children who had played CliniPup before the operation had significantly lower anxiety scores. Liu et al.\cite{LiuImpacts} considered prenatal mental state and physiological indicators of pregnant women. Pregnant women are not patients who have psychological pressure and physical discomfort before the operation, and they have to endure great pain. Their research asked pregnant women to use a video game that provides different hospital scenes to improve childbirth knowledge and help reduce postoperative depression. The studies show that SGMO can stabilize emotions, eliminate preoperative fears, and actively cooperate with the operation.

\subsubsection{Using SGMO during a medical operation}
The research in this area is in its infancy because most operations require patients to avoid body movements. As a supplement to a traditional operation, the main purpose of SGMO is to use new methods to solve old problems and show the functional idea. Intraoperative wake-up surgical resection is the latest strategy to detect whether the brain function area has been damaged. Dagmar Turner, a 53-year-old musician woman, was diagnosed with a brain tumor and required an operation to remove it. Because the nerves that control hand movement and coordination are involved in the operation, the doctor asked Turner to play the violin during the operation to ensure that the brain parts that control the hand's movement and coordination are not damaged. Buehler et al.\cite{2017Intraoperative} explored the  influence of external factors on postoperative recovery in inguinal hernia repair. They randomly divided the children into a control group and an experimental group. During the operation, the children wore earphones, while the experimental group listened to music, and the control group did not. The results showed that external factors like music during an operation may reduce the incidence of postoperative maladaptive behavior in children. Similarly, Rosalie et al.\cite{A2019Music} studied Music Interventions in Pediatric Surgery and verified that this method could help reduce  postoperative pain. Although no SGs were used in these operations, they all reflect the new ideas in the process, as a supplement to the traditional operation.

\begin{figure*}[!t]
	\centering 
	\includegraphics[scale=1]{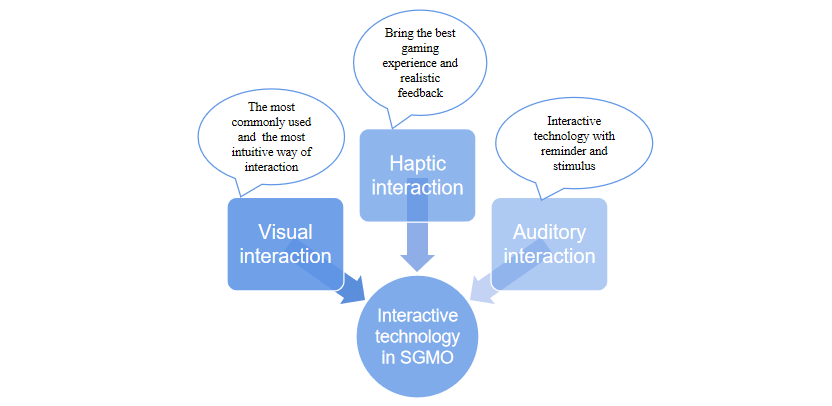}
	\caption{Interactive technologies used in SGMO and their characteristics}
	\label{fig1}
\end{figure*}

\subsubsection{Using Serious games after a medical operation}
After an operation, patients often suffer from postoperative pain and psychological pressure. Postoperative rehabilitation is a targeted treatment for patients with physical and psychological problems. Traditional rehabilitation includes medications, physical therapy, etc. SGMO for patients after operations is not a replacement for traditional rehabilitation but a supplementary way to carry out efficient rehabilitation training. Fernandez et al.\cite{7377254} used a game based on Kinect to help patients with rehabilitation training. The game’s three screens correspond to different training parts. When the player moves his limbs, the content changes accordingly. Michaud et al.\cite{8007482} created a game called The messy dock, which is  based on The Lokomat (a robotic treadmill training system) to help patients with gait training. Players can move their feet to cross obstacles in the game for rehabilitation. Heins et al.\cite{7939262} created a ROBiGAME game for sports and cognitive rehabilitation. This game can evaluate patients’ postoperative performance and improve the rehabilitation effect. Morrow et al.\cite{2006Low} provided an Xbox-based physical rehabilitation system in clinical rehabilitation, and the patient wore game gloves to complete some actions. The game can generate a butterfly, allowing the patient to touch the butterfly through hand movements to improve the fingers' flexibility and movement speed. Similarly, Alexandre et al.\cite{2019Physical} focused on the functional recovery of patients' limbs and fingers. The game is associated with gloves and headphones to collect various data through body movements. Ingadottir et al.\cite{ingadottir2017development} proved in experiments that SGMO was beneficial to patients' pain management after MOs. Bonnechère et al.\cite{bonnechere2016use} thought the therapy of SGMO is at least as effective as traditional therapies, and serious games are more popular with patients and should be integrated into various treatments. Lopes et al.\cite{S2018Games} confirmed that SGMO had a positive effect on postoperative rehabilitation and recommended them as a supplement to traditional therapy. In the postoperative recovery phase, SGMO also play an important role in helping them recover quickly.

\section{Application of interactive technologies in SGMO}
\label{sec:technology}
The medical staff need to pay attention to the patient's wound and related physiological indicators during an MO. This interaction between medical staff and patients involves multiple senses. Mack et al.\cite{mack2012interactive} pointed out that in an ideal simulated surgery or telerobotic surgery world, the medical staff should not feel the distance from the patient, all the information acquired by the senses should be realistic, and the most important senses are vision, touch, and hearing. In terms of technology, to enhance the game's authenticity and effectiveness, sensory interaction is often used to simulate the real operational experience. Some games focus on visual interaction, and other games focus on haptic interaction. However, in the existing literature, there is no research on the interactive technology from a  multi-sensory perspective. Therefore, in this section, we study the application of interactive technology in SGMO from three perspectives that are most closely with MOs: visual, haptic, and auditory senses. Their relationship is shown in Fig. 1.

\subsection{Visual interaction}
Medical staff can get  information from visual interaction, seize any details in the operating room, and communicate with others what they see during the MOs. Vision is the most commonly used and intuitive sense in SGMO, and many operations are based on visual observation. Board games in SGMO generally focus on acquiring and accumulating knowledge and do not require high visual interaction. Whittam et al.\cite{Alexander2017An} proposed a game called Burns to educate medical staff on evaluating and treating burns. The players play chess according to the game's rules, and the visual manifestation in the game is that they focus on the board set by the game, browse the problems, and interact with their peers. Graafland et al.\cite{graafland2015appraisal} developed a serious game that depicts a simulated laparoscopic tower, and the player should diagnose the problem. Some small icons are used to represent the patient's various physiological indicators, and some dynamic changes in graphics represent the event generated in the process. This game's visual interaction allows players to focus on the graphics changes on the screen, which can improve  situational awareness. IJgosse et al.\cite{2018Saving} created the Underground, which is an adventure game with 2D graphics. The visual effect of the participants is like watching a movie and experiencing the adventure  themselves. Kowalewski et al.\cite{2017Validation} created a 2D game named Touch Surgery™, which is different from the Underground, showing a fictional scene. It is the scene of a highly simulated cholecystectomy operation, and the player looks at the screen on the iPad as if looking at the scene on the operating table. Lange et al.\cite{2009Breath} used a game called Breath to help patients with postoperative breathing training. The game screen shows a bird, and the patient controls the bird's flight by exhaling and inhaling. This visual feedback could effectively help patients recover after MOs.\indent\par

The development of 3D technology makes the visual interaction of SGMO more sophisticated. Sliney et al.\cite{2008JDoc} proposed a 3D serious game called JDoc to simulate a doctor who can talk to the nurse and evaluate the patient's condition. The player can freely switch the perspective of the conversation in the game. Chon et al.\cite{2019Serious} used a 3D game called EMERGE, which shows a hospital scene. 3D modeling was more realistic, and the visual effects were better than JDoc. The players imagine that they are  an emergency doctor and go  to different departments to communicate with virtual patients and observe their physiological indicators. In the SGMO with 3D graphics, there are also many studies devoted to enhancing 3D visual effects. Sabri et al.\cite{sabri2010off} used shader rendering and created three textures to make 3D serious games more realistic. Considering the cutting operation during an MO, Hu et al.\cite{7960058} proposed a new type of 3D model dynamic cutting technology, using a texture supplement algorithm, which can enhance the visual authenticity in the simulation cutting of the 3D model. Lee et al.\cite{lee1999interactive} reconstructed and visualized the colon's inner surface. They generated high-resolution video views of the colon interior structures as if the eyes were inside the colon, which enhanced the visual experience.\indent\par

In addition to the game itself, some SGMO also added external devices to improve visual interaction. Jayakumar et al.\cite{7433879} simulated cataract surgery using Leap Motion equipment to capture the user's hand movements. The user can use gestures to interact with video content and use elements on the screen. The user's visual experience can be enhanced by hand movements to complete the cataract surgery's intraocular lens implantation process. Piedra et al.\cite{7590375} also paid attention to the enhancement of visualization by Leap Motion equipment during laparoscopic surgery. In the game environment, players use gestures to manipulate tools such as scissors, tweezers, and clips to simulate cutting and suture operations. The visual experience is similar to holding a scalpel for surgery. Carbone et al.\cite{carbone2016psychomotor} discussed the relationship between laparoscopic surgery and video game experience. They used the Xbox to control the game, and the graphics in the game can be moved by controlling the buttons of the Xbox, imitating the hand movements in laparoscopic surgery. The Underground\cite{2018Saving} also uses peripheral devices to enhance the visual experience. The player uses probes similar to laparoscopic instruments to rescue a virtual robot trapped in a mine. In terms of visual interaction, external devices use body movements as an input to control  changes in  the game's content, which enhances the visual effect and attracts the player’s attention.\indent\par

In recent years, virtual technology has developed rapidly and has also been widely used in SGMO. It includes: 1) Virtual reality, which uses computer simulation to generate a 3D virtual world, provides users with a simulation of vision and other senses, makes players feel immersed in the virtual world and can freely switch the observation perspective. 2) Augmented reality, which refers to the technology that allows the virtual world on the screen to be combined and interacted with the real world. These two technologies can give players an immersive experience where visual interaction has been greatly optimized.\indent\par

De Lima et al.\cite{2016A} used a virtual reality game named DOC TRAINING for clinical teaching. Players need to wear the GoogleCardboard virtual glasses to enjoy the immersive visual experience. In this game, players can use interactive visual technology to talk to simulated patients to diagnose their condition. Harrison et al.\cite{7939269} used a serious game called Through the Eye of the Master in surgical teaching. The player wore a Samsung Gear VR headset to feel immersive, and  virtual images were taken by a stereoscopic camera installed on the player’s head, allowing the player to watch the video from a first-person point of view. Cai et al.\cite{8949211} also used virtual games as clinical teaching for medical staff. The surgical environment is highly simulated, and the human body and surgical instruments are modeled in 3D, which is  convenient for players to observe the position of the organs in the human body. Players can also continuously switch the view and call a custom interface to observe the patient's organs.\indent\par

Grandi et al.\cite{7067098} used AR technology to observe human organs and tissues. The game system allows the visualization of the body's internal structures using an iPad as a mobile display. The camera captures markers on the real object and generates a corresponding picture on the screen as if there was no physical obstacle between them. Sato et al.\cite{9015418} used AR technology in the communication between doctors and patients to help doctors explain to patients their condition and the surgery simply and clearly. This game system aims at providing simple illustrations on paper media, using 3D models on portable systems. Larrarte and Alban\cite{7743367} explored visual interaction in mobile augmented reality. The game content simultaneously displays the real environment around the players and virtual objects in augmented reality, which can extract depth and marks from medical objects.\indent\par

\begin{figure*}[!t]
	\centering 
	\includegraphics[scale=1]{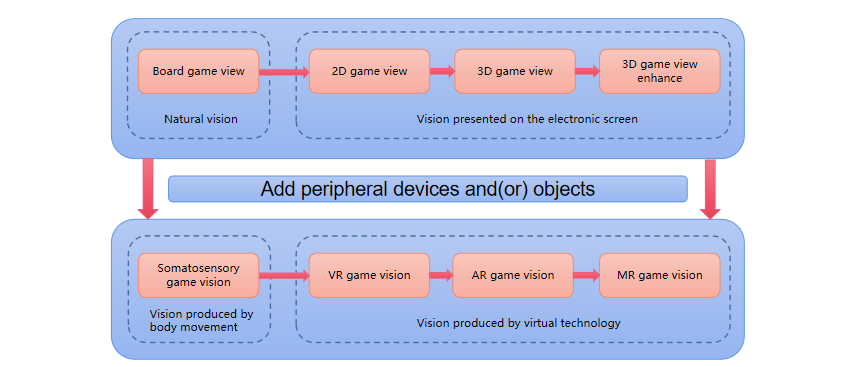}
	\caption{Innovation and characteristics of visual technology in SGMO}
	\label{fig2}
\end{figure*}

Panzoliet et al.\cite{8864546} studied visual interaction in MR. The experimental device is equipped with an HTC VIVE headset on which a Leap Motion and a webcam are  installed to perform hand detection and object detection, respectively. The user can see the virtual hand and virtual cubes and control them. Maasthi et al.\cite{9027491} used an application to study visual interaction in rhinoplasty surgery and used visual-tangible mixed reality in the study. The 3D face is imported to the game platform. Besides, unity animations and Mixed reality toolkit(MRTK) packages are applied. That allows the face model to be projected onto the holographic lens or the physical world. The medical staff and the patients can communicate based on the face model.\indent\par

A summary of the interactive visual technology and its technological innovation and characteristics is shown in Fig. 2.

\subsection{Haptic interaction}
The sensation produced by the skin's tactile receptors in contact with mechanical stimuli is called touch. Chen et al.\cite{662877} proposed that the medical staff interact with the patients through touch to  significantly enhance the experience of operation. Although some simulated surgeries allow players to interact with the screen content using a mouse and keyboard, this interaction only relies on vision to judge whether the player's operation is reasonable. As for the strength and depth of the cutting, no feedback can be obtained. Researchers believe that surgery simulators or serious games can bring a realistic visual experience, and if haptic sensations are provided, the experience will be significantly enhanced. In this paper, haptic interaction is used to describe the combination of kinesthetic and tactile feedback, including the perception of pressure, vibration, and texture.\indent\par

In the early exploration of force feedback, some surgical simulators appeared to train medical staff’s skills. For example, the PC-based MIST VR system from Virtual Presence is used to train doctors' skills in minimally invasive surgery, and so are the Virtual Clinic simulation from CineMed and the OR View by MusculoGraphics, Inc. Immersion Corporation provided force feedback equipment to integrate these systems, as well as some interfaces and engines. Pinzon et al.\cite{2016Prevailing} listed some simulators commonly used in surgery, such as the MIST VR, Lap Mentor II, andLapSim. MIST VR is a virtual simulator, which is initially set as a task simulator rather than a real environment simulation, so it lacks the mechanical characteristics. LapSim has real human tissue and tool interaction capabilities. The system displays various feedback measures, including speed, efficiency, and accuracy measurements. The Xitact IHP handles and instrument ports provide LapSim with haptic feedback that shows the friction was not inherent in laparoscopic surgery and limited the structure's effectiveness when using the haptic function. When interacting with tissues, The Lap Mentor uses microbot and software to create and improve haptic sense. State-of-the-art soft tissue interaction capabilities simulate the most complex anatomy, different tissue types and dissection planes, and provide realistic tool tissue interaction.\indent\par

\begin{table*}[]
	\centering
	\label{Table3}
	\setlength{\tabcolsep}{3pt} 
	\renewcommand\arraystretch{2.0} 
	\caption{HAPTIC SIMULATION METHODS AND CHARACTERISTICS IN MEDICAL OPERATION}
	\begin{tabular}{|c|c|c|c|}
		\hline
		Simulation project                                                                                       & Simulation method      & Advantages                                                                                                              & Disadvantages                                                                                     \\ \hline
		\multirow{3}{*}{\begin{tabular}[c]{@{}c@{}}Haptic\\  simulation\\  in Medical\\  Operation\end{tabular}} & Box-training(BT)       & \begin{tabular}[c]{@{}c@{}}Provides natural\\  tactile feedback\end{tabular}                                            & \begin{tabular}[c]{@{}c@{}}Lacks a realistic\\  environment and\\  objective results\end{tabular} \\ \cline{2-4} 
		& Virtual reality (VR)   & \begin{tabular}[c]{@{}c@{}}Provides objective\\  results\end{tabular}                                                   & \begin{tabular}[c]{@{}c@{}}Lacks haptic\\  feedback\end{tabular}                                  \\ \cline{2-4} 
		& Augmented reality (AR) & \begin{tabular}[c]{@{}c@{}}Provides natural\\  tactile, realistic\\  environment, and\\  objective results\end{tabular} & Not mentioned yet                                                                                 \\ \hline
	\end{tabular}
\end{table*}

There are other devices used for haptic interaction in surgical simulation. For patients with paralysis or neurological disorders, Brain-Computer Interface (BCI) can help them interact with the world. The P300 speller was a kind of BCI which enables paralyzed people to type directly using their brain signals. Choi et al.\cite{8241655} used a miniature vibrating motor to provide P300 with vibrating tactile force to improve BCI performance. Wischgoll et al.\cite{1372281} added a haptic device to the game to study coronary heart disease to enhance interaction. The input device uses SideWinder force-feedback wheel, and the force feedback signal is used to represent the collision. The force induced in the game device is proportional to the collision's force and speed between the camera and the blood vessel wall. The PHANToM device was also a device with a tactile interface. The Phantom Omni (Sensable Inc.)\cite{6193304} is  a 3-DOF haptic device that users hold in their hand just like a stylus. When moving the haptic device, the pointer moves accordingly in the 3D environment. The haptic 3D environment is derived from the Kinect data, and a  haptic algorithm successfully presents the haptic force of non-moving and mobile object interaction. The work\cite{6193304} provided ideas for tactile interaction in remote robotic operation. In addition to studying the application of artificial tactile devices, some researchers also focused on the application of natural tactile devices. Overtoom et al.\cite{overtoom2019haptic} proposed that Box Trainer (BT) provided natural tactile sensation, which can be used for surgical simulation, but the BT lacked real environment and objective results. Considering the combination of tactile and objective results, augmented reality training can be used in tactile simulators and provide a new direction for tactile research. The haptic simulation methods and characteristics in MO are listed in TABLE III.\indent\par

The equipment and products mentioned previously were used to research haptic interaction, combining functional ideas to offer a realistic environment to medical staff and improve their skills. Next, we specifically study haptic interaction in serious games.\indent\par

Toledo et al.\cite{2017Haptic} proposed a serious game in cataract surgery training based on  the Geomagic Touch equipment. Multi-layer eyeball models are built in the game, where each layer has a different texture. The structural parameters are adjusted according to factors such as stiffness and friction. When performing a simulated eyeball surgery, the user feels resistance during the micro-hole puncture operation, which makes the model enter a perforation state. Different perforation depths  produce different levels of resistance. Chan et al.\cite{6216416} used  a 6-DOF haptic device, PHANTOM Premium 1.5 High Force, to simulate the needle in serious games and let the doctor learn Ultrasound-guided needle placement skills. Qin et al.\cite{2010Learning} developed a serious game called Stopping the Fountains to learn Blood Management in Orthopedic Surgery. The hardware devices of the game include tactile input devices that manipulate virtual tools, present feedback during virtual operations, and use the haptic library API (HLAPI) in the OpenHaptics tool kit to simulate the touch feeling of contact between virtual instruments and tissues. Players create surgical incisions, stretch the wounds, fix them with hooks, and use virtual sealers to seal blood vessels. Throughout the game process, players can get tactile feedback to control the operation's accuracy, bringing immersive experience.\indent\par

Haptic interaction in remote operations has also been a hot research direction in recent years. In remote surgery, the surgical robot performs actions according to the doctor's remote instructions. However, doctors cannot get the force feedback exerted by the robot on the patient's body, which is a limitation. Mack et al.\cite{mack2012interactive} used PHANTOM Omnis with an attached instrument handle to study haptic interaction in remote operations. When the surgeon moves the handles of the surgical platform, the remote robot's arms also move with his actions. The integrated QTC force sensor collects force values during the operation, and doctors can log in to the VPRN client to connect with a remote server and obtain force feedback data. Then, the system presents these data as force feedback for the doctor to experience. Kulkarni et al.\cite{mack2012interactive} created a virtual animation environment in which the players wear tactile gloves to control the virtual hand model's movement, and the tactile feeling is transmitted through the tactile gloves. Munawar et al.\cite{7759188} used the DaVinci Research Kit (dVRK) to study the sense of touch in remote surgery. A real surgical manipulator was used in the research instead of tactile devices. At present, the research of haptic interaction in remote MOs is mainly based on some simulators and frameworks, and specific SGs have not been widely studied.\indent\par

\subsection{Auditory interaction}
Auditory interaction also has a wide range of applications in MOs. The medical staff need to pay attention to the live sound of the operating room and instructions of various instruments, and infer whether the patient's physiological indicators are normal.\indent\par

Sabri et al.\cite{sabri2010off} added background sound recorded to SGs during a real Off-pump coronary artery bypass surgery(OCPAB). Although the background sound did not have a dialogue between medical staff and patient, it contained the environmental sound of an OCPAB operation such as the anesthetic machine's sound. They planned to add the conversation sound of the medical staff diagnosis in the follow-up research as a kind of information output to indicate whether the performance is good. Wischgoll et al.\cite{1372281} added not only haptic interaction but also sound feedback to the surgical training of coronary heart disease. Sound effects using the DirectSound software accompany the signal to provide additional cues to enhance the game experience and improve the training effect.\indent\par

Some researchers worked on auditory interactive games for patients. Rossi et al.\cite{8401355} used a VR game named Imaginator to research the application of SGs in the treatment of sensory processing disorders (SPDs), including auditory stimulation. Correa et al.\cite{4362120} proposed an augmented reality game named GenVirtual to deal with learning disabilities. The therapist can flexibly place music elements in the game to create different scenarios for different patients. When the game starts, a music sequence is generated first, and then multiple virtual cubes are generated and light up with the music sequence while it is being played. The patient interacts with the virtual cube according to the notes he is hearing, and if the patient chooses the correct one, the system adds new notes to increase the difficulty of the game. Boothroyd et al.\cite{217403} developed a serious game for children with hearing loss and/or auditory processing disorders. The children chase the animated character, using auditory sense to identify the position and hiding place of the character. Kurniawati et al.\cite{8973353} also developed serious games for children with Special Educational Needs (SENs). These children generally have physical or mental disabilities, and it was difficult for them to receive the same education compared to students of the same age. One of these games is a VR game called Class VR. One mode of this game is called Finding the item heard, which provides sound but no images. In this mode, players are required to answer questions according to system prompts, and both correct and incorrect answers give feedback in different intonations. Game’s questions are aural and the instructions are completed by voice commands combined with intonation.\indent\par

The functions and technologies for  medical staff and patients in the serious game of auditory interaction are given in TABLE IV.\indent\par

\begin{table}[]
	\centering
	\label{Table4}
	\setlength{\tabcolsep}{3pt} 
	\renewcommand\arraystretch{2.0} 
	\caption{TECHNOLOGIES AND FUNCTIONS FOR DIFFERENT GROUPS IN SERIOUS GAMES OF AUDITORY INTERACTION}
	\begin{tabular}{|c|c|c|}
		\hline
		Player/User   & Technology                                                                                                                            & Functions                                                                                                                                                      \\ \hline
		Medical staff & \begin{tabular}[c]{@{}c@{}}Provides real sounds\\  and prompts in the\\  operating rooms\end{tabular}                                 & \begin{tabular}[c]{@{}c@{}}Helps players to familiarize\\  themselves with the operational\\  environment and complete the\\  simulated operation\end{tabular} \\ \hline
		Patients      & \begin{tabular}[c]{@{}c@{}}Provides sound\\  stimulation, which can\\  change dynamically with\\  the patient's response\end{tabular} & \begin{tabular}[c]{@{}c@{}}Helps patients with sensory\\  impairment to receive MOs\\  and rehabilitation\end{tabular}                                         \\ \hline
	\end{tabular}
\end{table}

\section{Future research directions in SGMO}
Because of the combination of function and technology, the coverage of SGMO has significantly been expanded, and it can be found in the perioperative period. SGMO have great development potential, but there are also some shortcomings worthy of further  research. In the following, we suggest directions for research.\indent\par

\begin{enumerate}
	\item In MOs multiple doctors and nurses cooperate and exchange useful information. Current SGMO are mainly stand-alone games and lack an interactive mechanism. In the game, players can only complete designated actions according to instructions, such as the game's content and prompt sounds. They cannot revise their thoughts and actions through others’ instructions. Future SGMO should allow multiple players to participate and perform their duties. Besides, they should allow  medical staff and patients to join the same game  where they can communicate to make professional explanations easy to understand. In addition, patients can also show their surgical expectations through intuitive models.
	
	\item SGMO can allow content that is not realistic. However, the game's operating process must comply with the corresponding MOs, and the steps must be reviewed by medical experts, which is also conducive to the close connection between computer science and medicine. In the SGMO's development stages, the most important factor is to ensure medical  rigor, and fully accept experts' review and clinical experiment verification. The experiment aims to verify whether the operation process strictly follows the MO’s process.
	
	\item Adding multi-sensory interaction technology to SGMO will provide a more realistic game experience. During the operation, both medical staff and patients can receive information through the senses and transmit effective information. SGMO, as a supplementary method for MOs, need to comprehensively produce useful information. Sensory interaction is not limited to one way; many senses need to cooperate to achieve the best effect of processing information.
	
	\item Remote MOs require the doctor to control the operating platform to input control instructions, and the sensing device can also automatically acquire the doctor's location information. The surgical machine platform can perform  operations according to the doctors' instructions and actions. Remote MOs must be based on various commands, human motion perception, and real-time transmission of information. SGMO can strengthen information acquisition, information transmission capabilities, and control hardware devices in remote locations through local games and get real-time feedback.
\end{enumerate}

\section{Conclusion}
SGMO have developed rapidly in recent years and also strengthened the close connection between computer science and medicine. Understanding the importance and development of SGMO helps researchers explore the application of SGs in medicine. Therefore, we have divided SGMO into four stages comprehensively judged according to the game's form, function, and technology. We cannot simply assume that the first SGMO in the early period are worse than the latest ones. The players need to use the games according to their needs instead of pursuing the latest games.\indent\par

Because of the different functional requirements of players using SGMO at different periods, we reviewed the SGMO from a functional perspective and compared the games’ characteristics. The functions of SGMO have become more and more diverse as the types of games increase. As the two major users of games, medical staff, and patients have different needs at different stages. These functions are accompanied by technologies that support users. The interactive technology in SGMO is close to sensory interaction during MOs and conveys useful information to users. We concentrated on interactive technology from a visual, haptic, and auditory perspective and listed the representative products. Besides, issues such as teamwork, game design, and remote MOs are important and deserve further study.

%
%
%

\bibliographystyle{IEEEtran}
\bibliography{reference}
\end{document}